\documentclass[
preprint,
superscriptaddress,
 amsmath,amssymb,
aps,
prl,
floatfix,
]{revtex4-2}
\usepackage{comment}
\usepackage{graphicx}
\usepackage{dcolumn}
\usepackage{bm}

\begin{document}

\preprint{APS/DFT+GOU}

\title{
Identification of the Mott insulating CDW state in 1T-TaS$_2$}

\author{Dongbin Shin}
\affiliation{Max Planck Institute for the Structure and Dynamics of Matter and Center for Free-Electron Laser Science, Luruper Chaussee 149, 22761, Hamburg, Germany
}
\author{Nicolas Tancogne-Dejean}
\affiliation{Max Planck Institute for the Structure and Dynamics of Matter and Center for Free-Electron Laser Science, Luruper Chaussee 149, 22761, Hamburg, Germany
}
\author{Jin Zhang}
\affiliation{Max Planck Institute for the Structure and Dynamics of Matter and Center for Free-Electron Laser Science, Luruper Chaussee 149, 22761, Hamburg, Germany
}
\author{Mahmut Sait Okyay}
\affiliation{Department of Physics, Ulsan National Institute of
Science and Technology (UNIST), UNIST-gil 50, Ulsan 44919, Korea
}
\author{Angel Rubio}
\thanks{angel.rubio@mpsd.mpg.de}
\affiliation{Max Planck Institute for the Structure and Dynamics of Matter and Center for Free-Electron Laser Science, Luruper Chaussee 149, 22761, Hamburg, Germany
}
\affiliation{Nano-Bio Spectroscopy Group, Departamento de Fsica de Materiales, Universidad del Pas Vasco, 20018 San Sebastian, Spain}
\affiliation{Center for Computational Quantum Physics (CCQ), The Flatiron Institute, 162 Fifth avenue, New York NY 10010
}

\author{Noejung Park}
\thanks{noejung@unist.ac.kr}
\affiliation{Department of Physics, Ulsan National Institute of
Science and Technology (UNIST), UNIST-gil 50, Ulsan 44919, Korea
}

\begin{abstract}
We investigate the low-temperature charge-density-wave (CDW) state of bulk TaS$_2$ with a fully self-consistent DFT+U approach, over which the controversy has remained unresolved regarding the out-of-plane metallic band.
By examining the innate structure of the Hubbard U potential, we reveal that the conventional use of atomic-orbital basis could seriously misevaluate the electron correlation in the CDW state. 
By adopting a generalized basis, covering the whole David star, we successfully reproduce the Mott insulating nature with the layer-by-layer antiferromagnetic order. 
Similar consideration should be applied for description of the electron correlation in molecular solid.
\end{abstract}

\maketitle

A first-principle description of interacting many-body states has been a central issue in many areas of solid state theories. 
While the density functional theory (DFT) provides a formal route for finding the ground state energy of a many-electrons system, the self-consistent density seeking procedure, as formulated by Kohn and Sham (KS), has been usually practiced with approximated local or semi-local functionals~{\cite{Perdew1983,Kohn1965,Perdew1981,Ceperley1980,Perdew1996a,Perdew1992}}.
Those practical functionals lack the derivative discontinuity in the energy density functional~\cite{Janak1978,Andrade2011,Chai2013}, and usually underestimates the fundamental band gap and overestimates charge transfer~{\cite{Anisimov1991,Shin2016,Shin2013}}. 
Among various complementation schemes~{\cite{Perdew1981,Heyd2003}}, the KS scheme combined with the Hubbard U potential, now known as DFT+U, has been most widely used particularly in the descriptions of the correlation-dominated insulating nature– the Mott insulating state~\cite{Anisimov1993,Lichtenstein1995,Cococcioni2005,Agapito2015,Tancogne-Dejean2018}. 
On the other hand, when the on-site potential is not fully repulsive, as in cases of various bad-metal phases, the statistical samplings of multiple occupations are essentially required, as treated in the framework of the dynamical mean field theory (DMFT)~\cite{Georges1996,Kotliar2004}. 

A four-decades-long debate has remained unresolved for the low temperature phase of bulk TaS$_2$. 
Whereas many experiments reported the insulating CDW phase~\cite{Wilson1975,Fazekas1980,Dardel1992,Sipos2008,Han2015,Hellmann2010,Hellmann2012,Martino2020}, first-principle calculations with DFT and DFT+U have noticed that the in-plane Mott states constitute the metallic band dispersion along the out-plane direction~\cite{Darancet2014, Ritschel2015, Lee2019,Lazar2015}.  
It has experimentally reported that, on the formation of the low-temperature Mott insulating phase of the commensurate CDW phase at T$<200$~K, the atomic structures are reconstructed into the David-star pattern, packed in the $\sqrt{13} \times \sqrt{13}$ in-plane super-lattice~{\cite{Sipos2008,Han2015,Hellmann2010,Hellmann2012,Martino2020}}.
The Mott insulating nature of the monolayer has been described with the first-principle calculations, and also the corresponding the two-dimensional (2D) Hubbard model Hamiltonian has been evaluated in the frame work of DMFT~\cite{Perfetti2005,Perfetti2006,Yoshioka2009}.
However, on stacking of the 2D Mott insulating layers in the bulk 1T-TaS$_2$, the calculation results with DFT, DFT+U and spin-frozen hybrid functional lose the separation between the lower Hubbard and upper Hubbard bands, resulting in a spin-unpolarized metallic dispersion along the out-of-plane direction~\cite{Darancet2014, Ritschel2015, Lee2019,Lazar2015}.
These theoretical results are clearly contrasting with experimentally observed insulating nature~\cite{Wilson1975,Fazekas1980,Dardel1992,Sipos2008,Han2015,Hellmann2010,Hellmann2012,Martino2020}.
Leaving behind these inconsistencies, variety of interests have been gathered on complex phases diagram of TaS$_2$~{\cite{Sipos2008,Han2015}}.
Also, controllable phase transitions have been pursued to utilize the phase diagram as a memory device~{\cite{Vaskivskyi2015, Yoshida2017, Yu2015,Hellmann2010, Zhang2019}}.

To resolve the inconsistency between the previous experiments and theories for the bulk 1T-TaS$_2$, here we reexamine the innate structure of the DFT+U~\cite{Darancet2014, Ritschel2015,Lee2019}.
In the standard formulation of the DFT+U, the total energy is expressed as a functional of both the electron density and the density matrix constructed in the correlated subspace $E_{DFT + U} = {E_{DFT}}[\rho (\mathbf {r})] + {E_U}[n]$ {\cite{Anisimov1993,Lichtenstein1995,Cococcioni2005}}. 
The averaged two-body energy is separated into the on-site Coulomb ($U$) and exchange energy ($J$), and  by subtracting the doubly counted mean-field energy from the DFT energy functional, the effective Hubbard U energy can be expressed in terms of $\bar{U}=U-J$, as follows:
\begin{equation}
{E_U} = \sum\limits_I {\frac{{{\bar U_I}}}{2}\sum\limits_{m,m',\sigma } {\left\{ {n_{m,m'}^{I,\sigma }{\delta _{m,m'}} - n_{m,m'}^{I,\sigma }n_{m',m}^{I,\sigma }} \right\}} },
\end{equation}
where $n_{m,m'}^{I,\sigma } = \sum\limits_{i,\mathbf{k},\sigma }\frac{1}{N_{\mathbf{k}}} {\left\langle {\psi _{i,\mathbf{k}}^\sigma \left| {\hat P_{m,m'}^I} \right|\psi _{i,\mathbf{k}}^\sigma } \right\rangle } $, and $\hat P_{m,m'}^I = \left| {\phi _m^I} \right\rangle \left\langle {\phi _{m'}^I} \right|$.
Above, $\psi_{i,\mathbf{k}}^{\sigma}$ indicates the KS orbital for the $i$-th band with the Bloch vector $\mathbf{k}$ and spin state $\sigma$. $\phi_m^I$ represents the atomic-orbital basis, for the $m$-th orbital of the $I$-th atom. $N_{\mathbf{k}}$ indicates number of sampled k-points in the Brillouin zone. 
The functional variation $\delta E_{DFT+U}/\delta \psi_{i,\mathbf{k}}^\sigma$ leads to one-body potential that depends on the electron density of the band states and the density matrix of the projected subspace {\cite{Anisimov1993,Lichtenstein1995,Cococcioni2005}}.

\begin{figure}
\includegraphics[width=0.5\textwidth]{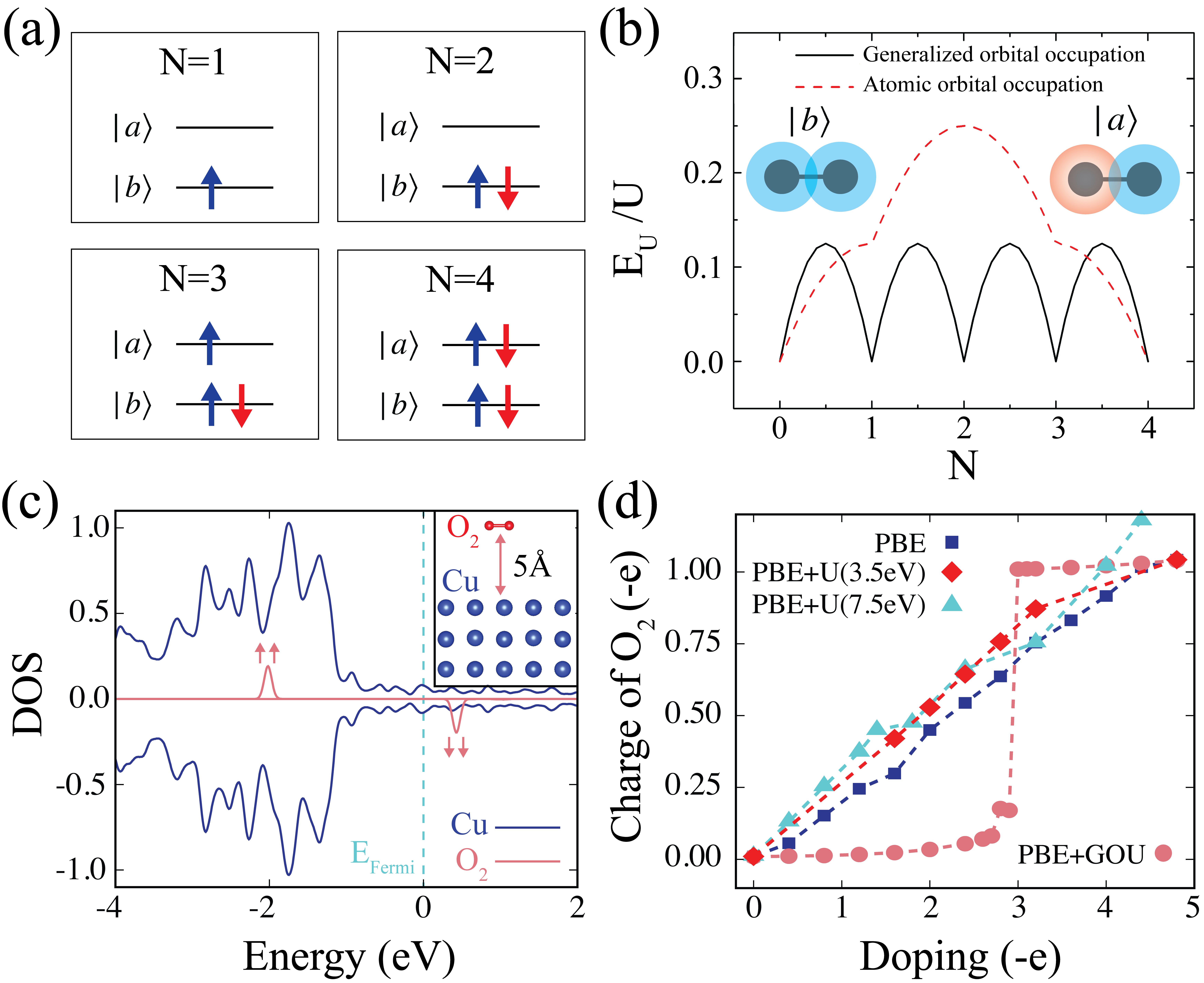}
\caption{
(a) Various occupations of the molecular orbitals of the model diatomic molecule. 
(b) The Hubbard energy with respect to the electron number (N) as defined in (a).
(c) Density of state for the O$_2$/Cu(100) system.
(d) The charge of O$_2$ with respect to additional electron doping on the O$_2$/Cu(100) system. 
In (a) and (b), $|a\rangle$ and $|b\rangle$ indicate the bonding and anti-bonding molecular orbitals, respectively. 
Inset of (c) indicates schematic geometry of O$_2$ on Cu(100) surface.
In (c), upward and downward arrows indicates doubly degenerated spin-up and spin-down states of O$_2$.
}
\end{figure}

Now we examine the characteristics of the functional with a simple molecular example. 
For a diatomic molecular orbital consisting of two one-orbital atoms, the Hubbard energy can be written as ${E_U} = \sum\limits_\sigma  {\frac{\bar U}{2}\left[ {\left( {{n_{b\sigma }} - n_{b\sigma }^2} \right) + \left( {{n_{a\sigma }} - n_{a\sigma }^2} \right)} \right]}$, where ${n_{b\sigma }}$ and ${n_{a\sigma }}$ represent the occupation numbers of the bonding and anti-bonding molecular orbitals with the spin state $\sigma$, respectively~\cite{Strebel1970,Chittipeddi1987,Lechermann2007}.
The sequence of electron occupations is schematically depicted in Fig.~1(a), assuming that the energy splitting between the bonding and anti-bonding states is substantially larger than the on-site interaction energy ($\bar{U}$).
The sharp kinks in the plot of $E_U$ in terms of electron numbers, as shown by the solid line in Fig.~1(b), illustrate clearly the well known derivative discontinuity of the exact exchange-correlation functional at the integer numbers of particles~{\cite{Janak1978, Andrade2011, Chai2013}}. 

One may consider the same system by counting the occupations of each atomic orbitals.
The Hubbard energy becomes ${E_U^{atom}} = \sum\limits_\sigma  {\frac{\bar U}{2}\left[ {\left( {{n_{1\sigma }} - n_{1\sigma }^2} \right) + \left( {{n_{2\sigma }} - n_{2\sigma }^2} \right)} \right]}$
with ${n_{1\sigma }}$ and ${n_{2\sigma }}$ indicating the occupation of the first and second atomic orbital with the spin ($\sigma$), respectively. 
Once the series of the electron occupation, shown in Fig. 1(a), is subject to the molecular symmetry, it is natural to claim that the two expressions of the density matrix are related by ${n_{1\sigma }} = {n_{2\sigma }} = \frac{1}{2}\left( {{n_{b\sigma }} + {n_{a\sigma }}} \right)$.
As revealed by the dashed red line in Fig.~1(b), the plot of the $E_U^{atom}$ in terms of the electron number largely misses the kink structure at the point of the integer numbers of electrons {\cite{Andrade2011, Cococcioni2005}}.
This result indicates that the application of the Hubbard U potential onto each constituent atomic orbitals are irrelevant, substantially misevaluating the electron correlation in molecular-orbital states.

To demonstrate this misbehavior in a more realistic example, we consider the O$_2$ molecule on the surface of Cu(100) surface: O$_2$ molecule fixed to a distance of 5\AA~above the three-layer Cu slab, as depicted in inset of Fig.~1(c). 
The density of state of the O$_2$/Cu(100) reveals the spin-triplet O$_2$ molecular state over the background metallic states of the Cu layer, as indicated by two pairs of arrows in Fig.~1(c).
We added additional electrons to this system, and we calculated each ground state with the given number of electrons, and the Bader charge is evaluated to account for the number of electron accommodated in O$_2$ as shown in Fig.~1(d). 
When we calculated the system by using the Perdew-Burke-Ehrenof (PBE)-type functional~\cite{Perdew1996}, these added electrons are fractionally distributed between O$_2$ and the metal layer, as shown in Fig.~1(d).
We now consider applying the U potential onto generalized basis, hereafter abbreviated as DFT+GOU, and it is applied on O$_2$ molecular orbital states for this example. 
The resulting number of electrons in O$_2$ exhibit a sharp step-like increase, which proves the discontinuous jump in the Coulomb energy of the electrons occupying the O$_2$ rather than long-range partial charge transfer.
Notably, the various U parameters in the conventional scheme resulted in similar fractional distributions, which clearly evidences our arguments on the misbehavior of the U applied onto the atomic orbitals.

In general, the U parameters are manually selected so as to reproduce the known electronic structure.
However, a few recent approaches has proposed schemes to determine the on-site Coulomb ($U$) and the exchange ($J$) parameters in a self-consistent way {\cite{Cococcioni2005,Kulik2011, Agapito2015, Mosey2008, Mosey2007,Tancogne-Dejean2018}}.
One of well-known methods was suggested by Agapito, Curtarolo, and Nardelli (ACBN0), in which the $U$ and $J$ parameters are determined through the theory of screened Hartree-Fock in the correlated subspace {\cite{Agapito2015,Tancogne-Dejean2019,Tancogne-Dejean2018}}. 
In our study, hereafter, the Coulomb parameter calculated by this self-consistent ACBN0 scheme is denoted by $\bar U$, while the manually selected one is noted by $U$.
For example, the computed parameter for O$_2$ molecular orbital in the O$_2$/Cu(100) system is found to be $\bar U \approx$ 7.5 eV. 
Key elements of the formulations of the DFT+GOU(ACBN0) are summarized and the revival of the derivative discontinuity are discussed in the Supplementary Material~\cite{SM}.

\begin{figure}
\includegraphics[width=0.5\textwidth]{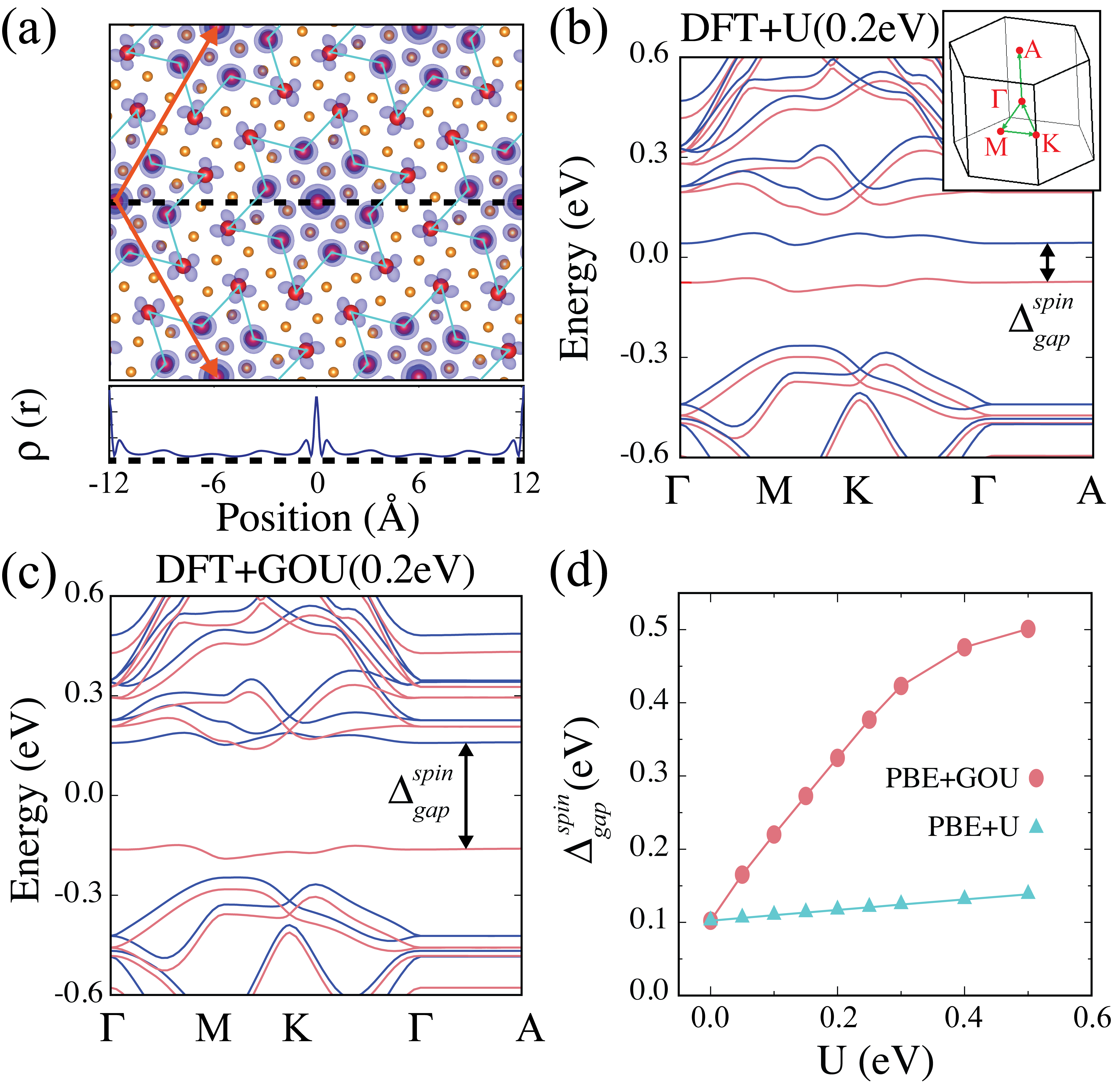}
\caption{
(a) An isosurface of the valence band maximum CDW state overlapped on the atomic geometry of the monolayer 1T-TaS$_2$ (upper panel), and the same charge density on the line that passes through the central Ta atom of the star-of-David (lower panel). 
Band structure of monolayer 1T-TaS$_2$ calculated by (b) DFT+U (${U}=0.2$eV) and (c) DFT+GOU (${U}=0.2$eV).
(d) The energy gap of monolayer 1T-TaS$_2$ with various Hubbard U values.
Two arrows in orange in the upper panel indicate the lattice vector of the $\sqrt{13}\times \sqrt{13}$ supercell. The lattice length is $a=12.13$~\AA. 
The line for the charge density plot is depicted by the thick dashed line in both panels. 
Red and orange ball indicate Ta and S atoms, respectively. 
The blue surface is isosurface of partial charge density of CDW state.
Cyan solid lines are the guide to the eye for David-star pattern of the CDW.
In (b) and (c), red and blue solid lines indicate spin-up and spin-down states, respectively.
The inset of (b) indicates the Bouillon zone of $\sqrt{13}\times \sqrt{13}$ supercell with high symmetric points.
}
\end{figure}

The aforementioned scheme is well suited for CDW state of the bulk 1T-TaS$_2$.
Figure~2(a) depicts the commensurate CDW state of the David star pattern in the $\sqrt{13}\times \sqrt{13}$ supercell. 
The lobe of charge density indicates that the CDW state mainly consists of $5d_z$ orbitals of the Ta atoms at the center and six surrounding vertex of the David star, which is also confirmed by the cross-sectional profile displayed in the lower panel of Fig.~2(a).  
The band structures of the monolayer calculated by DFT+U and DFT+GOU are displayed in Figs. 2(b) and 2(c), respectively, in which the flat band state below and above the Fermi level correspond to the singly occupied and unoccupied CDW states, respectively. 
For the generalized orbital to which the U potential is applied, we set the KS orbitals of the CDW state. 
Detailed procedures to derive the generalized basis orbital from the self-consistently converged Kohn-Sham states are summarized in the Supplementary Material~\cite{SM}.
The energy gap between the spin-up valence band and spin-down conduction band are summarized in Fig.~2(d).
The splitting between the upper and the lower Hubbard bands with respect to various U parameters obviously reveals the difference.
The energy gap calculated by the DFT+GOU increases linearly with the U parameter, while the spin gap size in the DFT+U is mostly inert over the increase of the U parameter.
Note that the difference in the Hubbard potential energy between occupied ($n_{CDW}=1$) and unoccupied CDW states ($n_{CDW}=0$) is exactly $U$, which can be read from $\hat{V}_{U} = \frac{1}{N_k}\sum\limits_{\sigma,\mathbf{k}} { U}(\frac{1}{2}-n_{CDW, \sigma}) | \phi_{CDW, \sigma, \mathbf{k}} \rangle  \langle \phi_{CDW, \sigma, \mathbf{k}} |$.

Now we focus on the band structure of bulk 1T-TaS$_2$, of which the out-of-plane conductivity has remained controversial for the last decade.
In our study, we chose the A-stacked bulk 1T-TaS$_2$ as a representative example, in which the Ta center atom of the David star is on the top of the same atom in the adjacent layer~\cite{Hovden2016, Darancet2014,Ritschel2015,Lee2019}.
We doubled the computational supercell to include two TaS$_2$ layers to investigate the layer-by-layer spin order.
The band structure of the bulk 1T-TaS$_2$ calculated by the DFT exhibit metallic band with a wide band width ($\sim 0.35$eV) along the out-plane direction ($\Gamma-$A), as shown in Fig.~3(a).
Furthermore, as depicted in the upper panel of Fig. 3(a), this metallic band doesn't produce the spin splitting, and the spin-up and spin-down charge densities exhibit almost the same pattern. 

\begin{figure}
\includegraphics[width=0.5\textwidth]{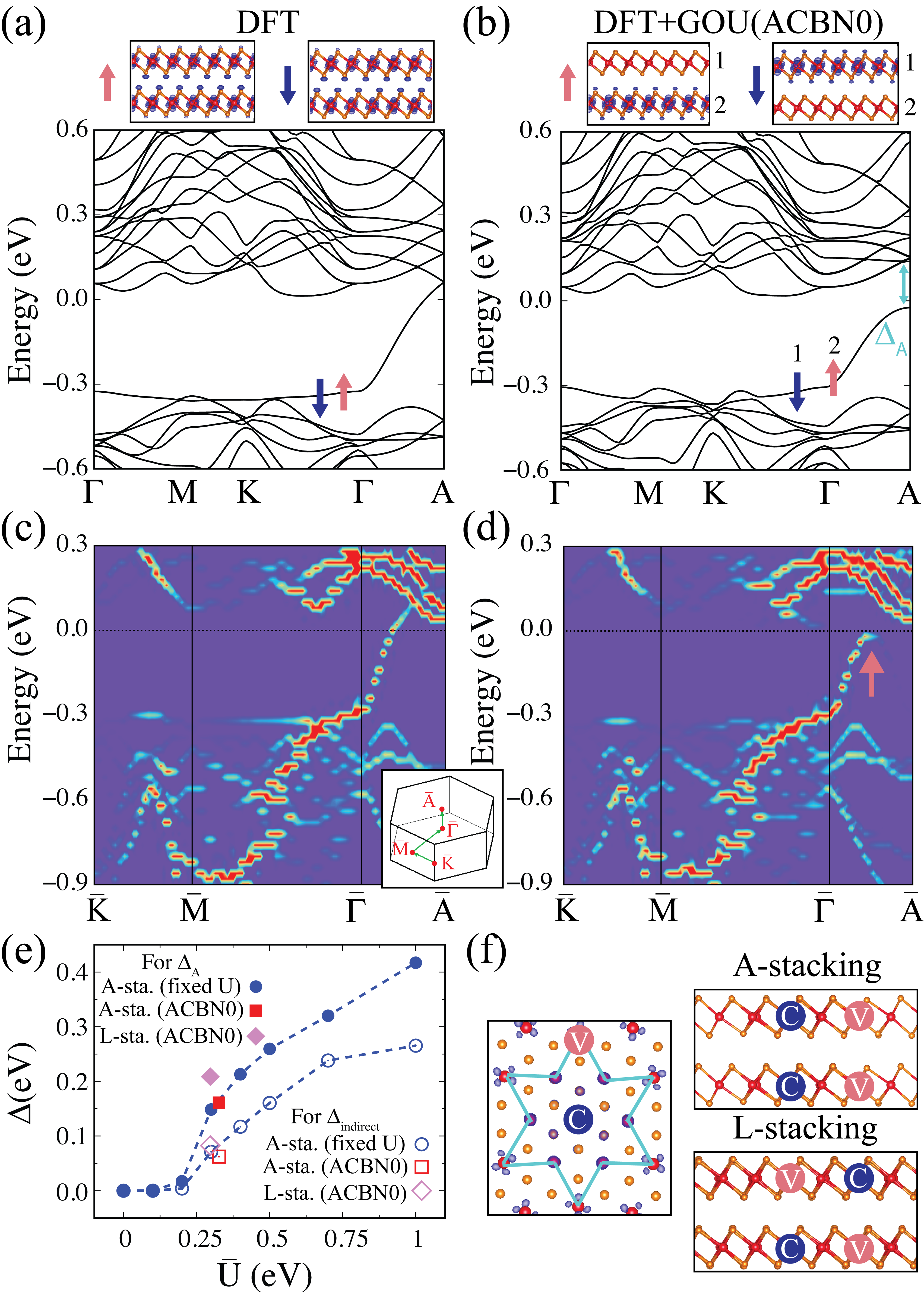}
\caption{
Band structure of the bulk 1T-TaS$_2$ calculated by (a) DFT and (b) DFT+GOU (ACBN0) methods.
Unfolded band structure of bulk 1T-TaS$_2$ in the primitive cell calculated by (c) DFT and (d) DFT+GOU (ACBN0) methods.
(e) The gap at zone boundary (A) and the indirect band gap of the bulk 1T-TaS$_2$ with respect to the various values of the U parameters.
(f) Schematic atomic geometry for A-stacking and L-stacking of the bulk 1T-TaS$_2$. 
In (a) and (b), the red and pink arrows indicate the spin-up and spin-down CDW states, respectively.
Upper panel of (a) and (b) indicate partial charge density of CDW state for spin-up (left) and spin-down (right) at zone boundary (the A point). 
The two adjacent layers are denoted by 1 and 2. 
The inset of (c) indicates the Bouillon zone of primitive cell with high symmetric points.
}
\end{figure}

On the other hand, the DFT+GOU scheme reveals an Mott insulating phase of the bulk 1T-TaS$_2$. 
In this study, the Hubbard U parameter is self-consistently determined by using the ACBN0 method ($\bar{U}=0.33$~eV), which produces a sizable gap opening: the energy gap at the zone boundary A point ($\Delta_{\text{A}}$) is $0.16$~eV and the indirect band gap ($\Delta_{\text{indirect}}$) is $0.06$~eV, which is eigen values difference between valence band maximum state at A point and conduction band minimum state at the point of 20\% along the path G-K. 
The experimental observed Mott gap ($\sim 0.12$~eV) is comparable with our computed values \cite{Hellmann2012}.
The CDW band exhibits clear spin splittings, and the spin-up and spin-down band edges reside in different layer as depicted by the partial charge densities of the CDW states shown in the upper panel of Fig. 3(b): the spin-down band in the first layer and the spin-up band in the next layer are denoted by the two numbered arrows. 
As a result, we obtained the layer-by-layer anti-ferromagnetic Mott insulating phase for the ground state of 1T-TaS$_2$.

For the comparison with experimental observations, we calculated the unfolded-band structure into the primitive cell of 1T-TaS$_2$ using the BandUP code~\cite{Medeiros2015}.
As shown in Figs.~3(c) and~3(d), the main difference between DFT and DFT+GOU methods can be found along the $\bar{\Gamma}$-$\bar{\text{A}}$ line.
The unfolded-band structure evaluated by the DFT+GOU(ACBN0) exhibits an apparent gap, which is in a good agreement with recent angle-resolved photoemission spectroscopy measurement in $\bar{\Gamma}$-$\bar{\text{A}}$ line~\cite{Ngankeu2017,Lee2019}.
For comparison, we tested with various U parameters, and the obtained energy gaps are summarized in Fig. 3(e): the gaps at the zone boundary at A point ($\Delta_{\text{A}}$) and the band gap indirectly formed at different k-points ($\Delta_{\text{indirect}}$) are displayed with solid and empty symbols, respectively. 

Now we consider different stacking orders and show that the obtained Mott insulating nature, in the result of DFT+GOU, is independent of the stacking order.
The center and vertex Ta atom of the David-star are denoted as C and V sites, as shown in the left panel of Fig. 4(f), respectively.
In the A-staking geometry, the center Ta atoms (C site) is exactly on top of the same C site in the adjacent layer, as depicted in the right upper panel of Fig.~3(f).
In the L-staking geometry, the C site of one layer is on top of the V site of the other layer, as shown in the right bottom panel of Fig. 3(f).
In our self-consistent DFT+GOU calculations with the ACBN0 functional, as summarized in Fig. 3(e), both the L-stacked and A-stacked bulk phases exhibit a similar size of the band gap. 

Numerous experimental studies have investigated the conductivity changes near the phase boundary of the bulk TaS$_2$~\cite{Wilson1975,Fazekas1980,Dardel1992,Sipos2008,Han2015,Hellmann2010,Hellmann2012,Martino2020}.
Near the phase transition (T$\sim200$~K), not only the in-plane conductivity but also the out-of-plane conductivity changes drastically~\cite{Martino2020}.
Measurements and discussions on the spin configurations are diverse~\cite{Perfetti2005,Perfetti2006,Law2017,Martino2020}, but the insulating nature of the bulk TaS$_2$ is experimentally agreed~\cite{Wilson1975,Fazekas1980,Dardel1992,Sipos2008,Martino2020}. 
A recent first principle calculations suggested that a very particular double-layer stacking order leads to a band gap without spin ordering~\cite{Ritschel2018,Lee2019}.
On the other hand, a recent scanning tunneling microscopy and a transmission electron microscopy revealed that the Ta atoms in different layers are in good order along the out-of-plane direction~\cite{Hovden2016}, and such type of double-layer order is not likely in practice~\cite{Butler2020}.
Our results of the DFT+GOU, as summarized in Fig. 3(e), indicates that the bulk preserves the Mott insulating phase irrespective of the stacking order, which supports all the experimental results mentioned above and suggest that the 3D insulating nature should obviously be attributed to the Coulomb correlation in the David star. 

In summary, we re-investigated the Mott insulating nature of bulk 1T-TaS$_2$ which electronic structure evaluated by DFT provides inconsistent results from previous experimental observations. 
We showed that the conventional Hubbard U potential can lead to erroneous evaluation of the electron correlation in KS states rooted in multiple atoms~\cite{Xian2019}. 
By applying the U potential on the generalized basis (DFT+GOU), we successfully reproduced the Mott insulating phases of the bulk 1T-TaS$_2$ with the layer-by-layer antiferromagnetic order, which resolves the decade inconsistent occurred by DFT calculation.
The electron correlation in orbitals localized, but extended over atoms, should be considered with similar perspective. 

$\it{Computational \ Details-}$
We performed DFT calculation using the Quantum Espresso package \cite{Giannozzi2017} with PBE-type functional~\cite{Perdew1996}.
The projector-augmented-wave method is used, and the plane wave basis set with 30 Ry and 60 Ry energy cut-off is used to describe wavefunction for 1T-TaS$_2$ and O$_2$/Cu(100) system, respectively.
The calculated lattice constant for bulk 1T-TaS$_2$ is $a=3.36$\AA~and $c=5.9$\AA~\cite{Givens1977}.
The Brillouin zone is sampled with $5 \times 5 \times 10$ $\mathbf{k}$-point mesh for 1T-TaS$_2$ systems, and the $8 \times 8 \times 1$ $\mathbf{k}$-point sampling is used for O$_2$/Cu(100) system.

\begin{acknowledgments}
We further acknowledge financial support from the European Research Council (ERC-2015-AdG-694097), the Cluster of Excellence 'CUI: Advanced Imaging of Matter' of the Deutsche Forschungsgemeinschaft (DFG) - EXC 2056 - project ID 390715994, Grupos Consolidados (IT1249-19), and SFB925 "Light induced dynamics and control of correlated quantum systems".
D.S. is supported by Alexander von Humboldt Foundation. 
N.P. is supported by National Research Foundation of Korea (NRF-2019R1A2C2089332).
J.Z. acknowledges funding received from the European Union Horizon 2020 research and innovation program under Marie Sklodowska-Curie Grant Agreement 886291 (PeSD-NeSL).
\end{acknowledgments}

\providecommand{\noopsort}[1]{}\providecommand{\singleletter}[1]{#1}%


\begin{thebibliography}{56}%
\makeatletter
\providecommand \@ifxundefined [1]{%
 \@ifx{#1\undefined}
}%
\providecommand \@ifnum [1]{%
 \ifnum #1\expandafter \@firstoftwo
 \else \expandafter \@secondoftwo
 \fi
}%
\providecommand \@ifx [1]{%
 \ifx #1\expandafter \@firstoftwo
 \else \expandafter \@secondoftwo
 \fi
}%
\providecommand \natexlab [1]{#1}%
\providecommand \enquote  [1]{``#1''}%
\providecommand \bibnamefont  [1]{#1}%
\providecommand \bibfnamefont [1]{#1}%
\providecommand \citenamefont [1]{#1}%
\providecommand \href@noop [0]{\@secondoftwo}%
\providecommand \href [0]{\begingroup \@sanitize@url \@href}%
\providecommand \@href[1]{\@@startlink{#1}\@@href}%
\providecommand \@@href[1]{\endgroup#1\@@endlink}%
\providecommand \@sanitize@url [0]{\catcode `\\12\catcode `\$12\catcode
  `\&12\catcode `\#12\catcode `\^12\catcode `\_12\catcode `\%12\relax}%
\providecommand \@@startlink[1]{}%
\providecommand \@@endlink[0]{}%
\providecommand \url  [0]{\begingroup\@sanitize@url \@url }%
\providecommand \@url [1]{\endgroup\@href {#1}{\urlprefix }}%
\providecommand \urlprefix  [0]{URL }%
\providecommand \Eprint [0]{\href }%
\providecommand \doibase [0]{https://doi.org/}%
\providecommand \selectlanguage [0]{\@gobble}%
\providecommand \bibinfo  [0]{\@secondoftwo}%
\providecommand \bibfield  [0]{\@secondoftwo}%
\providecommand \translation [1]{[#1]}%
\providecommand \BibitemOpen [0]{}%
\providecommand \bibitemStop [0]{}%
\providecommand \bibitemNoStop [0]{.\EOS\space}%
\providecommand \EOS [0]{\spacefactor3000\relax}%
\providecommand \BibitemShut  [1]{\csname bibitem#1\endcsname}%
\let\auto@bib@innerbib\@empty
\bibitem [{\citenamefont {Perdew}\ and\ \citenamefont
  {Levy}(1983)}]{Perdew1983}%
  \BibitemOpen
  \bibfield  {author} {\bibinfo {author} {\bibfnamefont {J.~P.}\ \bibnamefont
  {Perdew}}\ and\ \bibinfo {author} {\bibfnamefont {M.}~\bibnamefont {Levy}},\
  }\href@noop {} {\bibfield  {journal} {\bibinfo  {journal} {Phys. Rev. Lett.}\
  }\textbf {\bibinfo {volume} {51}},\ \bibinfo {pages} {1884} (\bibinfo {year}
  {1983})}\BibitemShut {NoStop}%
\bibitem [{\citenamefont {Kohn}\ and\ \citenamefont {Sham}(1965)}]{Kohn1965}%
  \BibitemOpen
  \bibfield  {author} {\bibinfo {author} {\bibfnamefont {W.}~\bibnamefont
  {Kohn}}\ and\ \bibinfo {author} {\bibfnamefont {L.~J.}\ \bibnamefont
  {Sham}},\ }\href@noop {} {\bibfield  {journal} {\bibinfo  {journal} {Phys.
  Rev.}\ }\textbf {\bibinfo {volume} {140}},\ \bibinfo {pages} {A1133}
  (\bibinfo {year} {1965})}\BibitemShut {NoStop}%
\bibitem [{\citenamefont {Perdew}\ and\ \citenamefont
  {Zunger}(1981)}]{Perdew1981}%
  \BibitemOpen
  \bibfield  {author} {\bibinfo {author} {\bibfnamefont {J.~P.}\ \bibnamefont
  {Perdew}}\ and\ \bibinfo {author} {\bibfnamefont {A.}~\bibnamefont
  {Zunger}},\ }\href@noop {} {\bibfield  {journal} {\bibinfo  {journal} {Phys.
  Rev. B}\ }\textbf {\bibinfo {volume} {23}},\ \bibinfo {pages} {5048}
  (\bibinfo {year} {1981})}\BibitemShut {NoStop}%
\bibitem [{\citenamefont {Ceperley}\ and\ \citenamefont
  {Alder}(1980)}]{Ceperley1980}%
  \BibitemOpen
  \bibfield  {author} {\bibinfo {author} {\bibfnamefont {D.~M.}\ \bibnamefont
  {Ceperley}}\ and\ \bibinfo {author} {\bibfnamefont {B.~J.}\ \bibnamefont
  {Alder}},\ }\href@noop {} {\bibfield  {journal} {\bibinfo  {journal} {Phys.
  Rev. Lett.}\ }\textbf {\bibinfo {volume} {45}},\ \bibinfo {pages} {566}
  (\bibinfo {year} {1980})}\BibitemShut {NoStop}%
\bibitem [{\citenamefont {Perdew}\ \emph
  {et~al.}(1996{\natexlab{a}})\citenamefont {Perdew}, \citenamefont {Burke},\
  and\ \citenamefont {Ernzerhof}}]{Perdew1996a}%
  \BibitemOpen
  \bibfield  {author} {\bibinfo {author} {\bibfnamefont {J.~P.}\ \bibnamefont
  {Perdew}}, \bibinfo {author} {\bibfnamefont {K.}~\bibnamefont {Burke}},\ and\
  \bibinfo {author} {\bibfnamefont {M.}~\bibnamefont {Ernzerhof}},\ }\href@noop
  {} {\bibfield  {journal} {\bibinfo  {journal} {Phys. Rev. Lett.}\ }\textbf
  {\bibinfo {volume} {77}},\ \bibinfo {pages} {3865} (\bibinfo {year}
  {1996}{\natexlab{a}})}\BibitemShut {NoStop}%
\bibitem [{\citenamefont {Perdew}\ and\ \citenamefont
  {Wang}(1992)}]{Perdew1992}%
  \BibitemOpen
  \bibfield  {author} {\bibinfo {author} {\bibfnamefont {J.~P.}\ \bibnamefont
  {Perdew}}\ and\ \bibinfo {author} {\bibfnamefont {Y.}~\bibnamefont {Wang}},\
  }\href@noop {} {\bibfield  {journal} {\bibinfo  {journal} {Phys. Rev. B}\
  }\textbf {\bibinfo {volume} {45}},\ \bibinfo {pages} {13244} (\bibinfo {year}
  {1992})}\BibitemShut {NoStop}%
\bibitem [{\citenamefont {Janak}(1978)}]{Janak1978}%
  \BibitemOpen
  \bibfield  {author} {\bibinfo {author} {\bibfnamefont {J.~F.}\ \bibnamefont
  {Janak}},\ }\href@noop {} {\bibfield  {journal} {\bibinfo  {journal} {Phys.
  Rev. B}\ }\textbf {\bibinfo {volume} {18}},\ \bibinfo {pages} {7165}
  (\bibinfo {year} {1978})}\BibitemShut {NoStop}%
\bibitem [{\citenamefont {Andrade}\ and\ \citenamefont
  {Aspuru-Guzik}(2011)}]{Andrade2011}%
  \BibitemOpen
  \bibfield  {author} {\bibinfo {author} {\bibfnamefont {X.}~\bibnamefont
  {Andrade}}\ and\ \bibinfo {author} {\bibfnamefont {A.}~\bibnamefont
  {Aspuru-Guzik}},\ }\href@noop {} {\bibfield  {journal} {\bibinfo  {journal}
  {Phys. Rev. Lett.}\ }\textbf {\bibinfo {volume} {107}},\ \bibinfo {pages}
  {183002} (\bibinfo {year} {2011})}\BibitemShut {NoStop}%
\bibitem [{\citenamefont {Chai}\ and\ \citenamefont {Chen}(2013)}]{Chai2013}%
  \BibitemOpen
  \bibfield  {author} {\bibinfo {author} {\bibfnamefont {J.~D.}\ \bibnamefont
  {Chai}}\ and\ \bibinfo {author} {\bibfnamefont {P.~T.}\ \bibnamefont
  {Chen}},\ }\href@noop {} {\bibfield  {journal} {\bibinfo  {journal} {Phys.
  Rev. Lett.}\ }\textbf {\bibinfo {volume} {110}},\ \bibinfo {pages} {033002}
  (\bibinfo {year} {2013})}\BibitemShut {NoStop}%
\bibitem [{\citenamefont {Anisimov}\ \emph {et~al.}(1991)\citenamefont
  {Anisimov}, \citenamefont {Zaanen},\ and\ \citenamefont
  {Andersen}}]{Anisimov1991}%
  \BibitemOpen
  \bibfield  {author} {\bibinfo {author} {\bibfnamefont {V.~I.}\ \bibnamefont
  {Anisimov}}, \bibinfo {author} {\bibfnamefont {J.}~\bibnamefont {Zaanen}},\
  and\ \bibinfo {author} {\bibfnamefont {O.~K.}\ \bibnamefont {Andersen}},\
  }\href@noop {} {\bibfield  {journal} {\bibinfo  {journal} {Phys. Rev. B}\
  }\textbf {\bibinfo {volume} {44}},\ \bibinfo {pages} {943} (\bibinfo {year}
  {1991})}\BibitemShut {NoStop}%
\bibitem [{\citenamefont {Shin}\ \emph {et~al.}(2016)\citenamefont {Shin},
  \citenamefont {Lee}, \citenamefont {Miyamoto},\ and\ \citenamefont
  {Park}}]{Shin2016}%
  \BibitemOpen
  \bibfield  {author} {\bibinfo {author} {\bibfnamefont {D.}~\bibnamefont
  {Shin}}, \bibinfo {author} {\bibfnamefont {G.}~\bibnamefont {Lee}}, \bibinfo
  {author} {\bibfnamefont {Y.}~\bibnamefont {Miyamoto}},\ and\ \bibinfo
  {author} {\bibfnamefont {N.}~\bibnamefont {Park}},\ }\href@noop {} {\bibfield
   {journal} {\bibinfo  {journal} {J. Chem. Theory Comput.}\ }\textbf {\bibinfo
  {volume} {12}},\ \bibinfo {pages} {201} (\bibinfo {year} {2016})}\BibitemShut
  {NoStop}%
\bibitem [{\citenamefont {Shin}\ \emph {et~al.}(2013)\citenamefont {Shin},
  \citenamefont {Jung}, \citenamefont {Han}, \citenamefont {Choi},
  \citenamefont {Lee},\ and\ \citenamefont {Park}}]{Shin2013}%
  \BibitemOpen
  \bibfield  {author} {\bibinfo {author} {\bibfnamefont {D.}~\bibnamefont
  {Shin}}, \bibinfo {author} {\bibfnamefont {H.}~\bibnamefont {Jung}}, \bibinfo
  {author} {\bibfnamefont {S.~S.}\ \bibnamefont {Han}}, \bibinfo {author}
  {\bibfnamefont {C.~H.}\ \bibnamefont {Choi}}, \bibinfo {author}
  {\bibfnamefont {H.}~\bibnamefont {Lee}},\ and\ \bibinfo {author}
  {\bibfnamefont {N.}~\bibnamefont {Park}},\ }\href@noop {} {\bibfield
  {journal} {\bibinfo  {journal} {Chem. Phys. Lett.}\ }\textbf {\bibinfo
  {volume} {570}},\ \bibinfo {pages} {85} (\bibinfo {year} {2013})}\BibitemShut
  {NoStop}%
\bibitem [{\citenamefont {Heyd}\ \emph {et~al.}(2003)\citenamefont {Heyd},
  \citenamefont {Scuseria},\ and\ \citenamefont {Ernzerhof}}]{Heyd2003}%
  \BibitemOpen
  \bibfield  {author} {\bibinfo {author} {\bibfnamefont {J.}~\bibnamefont
  {Heyd}}, \bibinfo {author} {\bibfnamefont {G.~E.}\ \bibnamefont {Scuseria}},\
  and\ \bibinfo {author} {\bibfnamefont {M.}~\bibnamefont {Ernzerhof}},\
  }\href@noop {} {\bibfield  {journal} {\bibinfo  {journal} {J. Chem. Phys.}\
  }\textbf {\bibinfo {volume} {118}},\ \bibinfo {pages} {8207} (\bibinfo {year}
  {2003})}\BibitemShut {NoStop}%
\bibitem [{\citenamefont {Anisimov}\ \emph {et~al.}(1993)\citenamefont
  {Anisimov}, \citenamefont {Solovyev}, \citenamefont {Korotin}, \citenamefont
  {Czyzyk},\ and\ \citenamefont {Sawatzky}}]{Anisimov1993}%
  \BibitemOpen
  \bibfield  {author} {\bibinfo {author} {\bibfnamefont {V.~I.}\ \bibnamefont
  {Anisimov}}, \bibinfo {author} {\bibfnamefont {I.~V.}\ \bibnamefont
  {Solovyev}}, \bibinfo {author} {\bibfnamefont {M.~A.}\ \bibnamefont
  {Korotin}}, \bibinfo {author} {\bibfnamefont {M.~T.}\ \bibnamefont {Czyzyk}},\
  and\ \bibinfo {author} {\bibfnamefont {G.~A.}\ \bibnamefont {Sawatzky}},\
  }\href@noop {} {\bibfield  {journal} {\bibinfo  {journal} {Phys. Rev. B}\
  }\textbf {\bibinfo {volume} {48}},\ \bibinfo {pages} {16929} (\bibinfo {year}
  {1993})}\BibitemShut {NoStop}%
\bibitem [{\citenamefont {Lichtenstein}\ \emph {et~al.}(1995)\citenamefont
  {Liechtenstein}, \citenamefont {Anisimov},\ and\ \citenamefont
  {Zaanen}}]{Lichtenstein1995}%
  \BibitemOpen
  \bibfield  {author} {\bibinfo {author} {\bibfnamefont {A.~I.}\ \bibnamefont
  {Liechtenstein}}, \bibinfo {author} {\bibfnamefont {V.I.}~\bibnamefont
  {Anisimov}},\ and\ \bibinfo {author} {\bibfnamefont {J.}~\bibnamefont
  {Zaanen}},\ }\href@noop {} {\bibfield  {journal} {\bibinfo  {journal} {Phys.
  Rev. B}\ }\textbf {\bibinfo {volume} {52}},\ \bibinfo {pages} {R5467}
  (\bibinfo {year} {1995})}\BibitemShut {NoStop}%
\bibitem [{\citenamefont {Cococcioni}\ and\ \citenamefont {{de
  Gironcoli}}(2005)}]{Cococcioni2005}%
  \BibitemOpen
  \bibfield  {author} {\bibinfo {author} {\bibfnamefont {M.}~\bibnamefont
  {Cococcioni}}\ and\ \bibinfo {author} {\bibfnamefont {S.}~\bibnamefont {{de
  Gironcoli}}},\ }\href@noop {} {\bibfield  {journal} {\bibinfo  {journal}
  {Phys. Rev. B}\ }\textbf {\bibinfo {volume} {71}},\ \bibinfo {pages} {035105}
  (\bibinfo {year} {2005})}\BibitemShut {NoStop}%
\bibitem [{\citenamefont {Agapito}\ \emph {et~al.}(2015)\citenamefont
  {Agapito}, \citenamefont {Curtarolo},\ and\ \citenamefont
  {Nardelli}}]{Agapito2015}%
  \BibitemOpen
  \bibfield  {author} {\bibinfo {author} {\bibfnamefont {L.~A.}\ \bibnamefont
  {Agapito}}, \bibinfo {author} {\bibfnamefont {S.}~\bibnamefont {Curtarolo}},\
  and\ \bibinfo {author} {\bibfnamefont {M.}~\bibnamefont {BuongiornoNardelli}},\
  }\href@noop {} {\bibfield  {journal} {\bibinfo  {journal} {Phys. Rev. X}\
  }\textbf {\bibinfo {volume} {5}},\ \bibinfo {pages} {011006} (\bibinfo {year}
  {2015})}\BibitemShut {NoStop}%
\bibitem [{\citenamefont {Tancogne-Dejean}\ \emph {et~al.}(2018)\citenamefont
  {Tancogne-Dejean}, \citenamefont {Sentef},\ and\ \citenamefont
  {Rubio}}]{Tancogne-Dejean2018}%
  \BibitemOpen
  \bibfield  {author} {\bibinfo {author} {\bibfnamefont {N.}~\bibnamefont
  {Tancogne-Dejean}}, \bibinfo {author} {\bibfnamefont {M.~A.}\ \bibnamefont
  {Sentef}},\ and\ \bibinfo {author} {\bibfnamefont {A.}~\bibnamefont
  {Rubio}},\ }\href@noop {} {\bibfield  {journal} {\bibinfo  {journal} {Phys.
  Rev. Lett.}\ }\textbf {\bibinfo {volume} {121}},\ \bibinfo {pages} {97402}
  (\bibinfo {year} {2018})}\BibitemShut {NoStop}%
\bibitem [{\citenamefont {Georges}\ \emph {et~al.}(1996)\citenamefont
  {Georges}, \citenamefont {Kotliar}, \citenamefont {Krauth},\ and\
  \citenamefont {Rozenberg}}]{Georges1996}%
  \BibitemOpen
  \bibfield  {author} {\bibinfo {author} {\bibfnamefont {A.}~\bibnamefont
  {Georges}}, \bibinfo {author} {\bibfnamefont {G.}~\bibnamefont {Kotliar}},
  \bibinfo {author} {\bibfnamefont {W.}~\bibnamefont {Krauth}},\ and\ \bibinfo
  {author} {\bibfnamefont {M.~J.}\ \bibnamefont {Rozenberg}},\ }\href@noop {}
  {\bibfield  {journal} {\bibinfo  {journal} {Rev. Mod. Phys.}\ }\textbf
  {\bibinfo {volume} {68}},\ \bibinfo {pages} {13} (\bibinfo {year}
  {1996})}\BibitemShut {NoStop}%
\bibitem [{\citenamefont {Kotliar}\ and\ \citenamefont
  {Vollhardt}(2004)}]{Kotliar2004}%
  \BibitemOpen
  \bibfield  {author} {\bibinfo {author} {\bibfnamefont {G.}~\bibnamefont
  {Kotliar}}\ and\ \bibinfo {author} {\bibfnamefont {D.}~\bibnamefont
  {Vollhardt}},\ }\href@noop {} {\bibfield  {journal} {\bibinfo  {journal}
  {Phys. Today}\ }\textbf {\bibinfo {volume} {57}},\ \bibinfo {pages} {53}
  (\bibinfo {year} {2004})}\BibitemShut {NoStop}%
\bibitem [{\citenamefont {Wilson}\ \emph {et~al.}(1975)\citenamefont {Wilson},
  \citenamefont {{Di Salvo}},\ and\ \citenamefont {Mahajan}}]{Wilson1975}%
  \BibitemOpen
  \bibfield  {author} {\bibinfo {author} {\bibfnamefont {J.~A.}\ \bibnamefont
  {Wilson}}, \bibinfo {author} {\bibfnamefont {F.~J.}\ \bibnamefont {{Di
  Salvo}}},\ and\ \bibinfo {author} {\bibfnamefont {S.}~\bibnamefont
  {Mahajan}},\ }\href@noop {} {\bibfield  {journal} {\bibinfo  {journal} {Adv.
  Phys.}\ }\textbf {\bibinfo {volume} {24}},\ \bibinfo {pages} {117} (\bibinfo
  {year} {1975})}\BibitemShut {NoStop}%
\bibitem [{\citenamefont {Fazekas}\ and\ \citenamefont
  {Tosatti}(1980)}]{Fazekas1980}%
  \BibitemOpen
  \bibfield  {author} {\bibinfo {author} {\bibfnamefont {P.}~\bibnamefont
  {Fazekas}}\ and\ \bibinfo {author} {\bibfnamefont {E.}~\bibnamefont
  {Tosatti}},\ }\href@noop {} {\bibfield  {journal} {\bibinfo  {journal}
  {Physica B+C}\ }\textbf {\bibinfo {volume} {99}},\ \bibinfo {pages} {183}
  (\bibinfo {year} {1980})}\BibitemShut {NoStop}%
\bibitem [{\citenamefont {Dardel}\ \emph {et~al.}(1992)\citenamefont {Dardel},
  \citenamefont {Grioni}, \citenamefont {Malterre}, \citenamefont {Weibel},
  \citenamefont {Baer},\ and\ \citenamefont {L{\'{e}}vy}}]{Dardel1992}%
  \BibitemOpen
  \bibfield  {author} {\bibinfo {author} {\bibfnamefont {B.}~\bibnamefont
  {Dardel}}, \bibinfo {author} {\bibfnamefont {M.}~\bibnamefont {Grioni}},
  \bibinfo {author} {\bibfnamefont {D.}~\bibnamefont {Malterre}}, \bibinfo
  {author} {\bibfnamefont {P.}~\bibnamefont {Weibel}}, \bibinfo {author}
  {\bibfnamefont {Y.}~\bibnamefont {Baer}},\ and\ \bibinfo {author}
  {\bibfnamefont {F.}~\bibnamefont {L{\'{e}}vy}},\ }\href@noop {} {\bibfield
  {journal} {\bibinfo  {journal} {Phys. Rev. B}\ }\textbf {\bibinfo {volume}
  {45}},\ \bibinfo {pages} {1462} (\bibinfo {year} {1992})}\BibitemShut
  {NoStop}%
\bibitem [{\citenamefont {Sipos}\ \emph {et~al.}(2008)\citenamefont {Sipos},
  \citenamefont {Kusmartseva}, \citenamefont {Akrap}, \citenamefont {Berger},
  \citenamefont {Forr{\'{o}}},\ and\ \citenamefont {Tutǐ}}]{Sipos2008}%
  \BibitemOpen
  \bibfield  {author} {\bibinfo {author} {\bibfnamefont {B.}~\bibnamefont
  {Sipos}}, \bibinfo {author} {\bibfnamefont {A.~F.}\ \bibnamefont
  {Kusmartseva}}, \bibinfo {author} {\bibfnamefont {A.}~\bibnamefont {Akrap}},
  \bibinfo {author} {\bibfnamefont {H.}~\bibnamefont {Berger}}, \bibinfo
  {author} {\bibfnamefont {L.}~\bibnamefont {Forr{\'{o}}}},\ and\ \bibinfo
  {author} {\bibfnamefont {E.}~\bibnamefont {Tutǐ}},\ }\href@noop {}
  {\bibfield  {journal} {\bibinfo  {journal} {Nat. Mater.}\ }\textbf {\bibinfo
  {volume} {7}},\ \bibinfo {pages} {960} (\bibinfo {year} {2008})}\BibitemShut
  {NoStop}%
\bibitem [{\citenamefont {Han}\ \emph {et~al.}(2015)\citenamefont {Han},
  \citenamefont {Zhou}, \citenamefont {Malliakas}, \citenamefont {Duxbury},
  \citenamefont {Mahanti}, \citenamefont {Kanatzidis},\ and\ \citenamefont
  {Ruan}}]{Han2015}%
  \BibitemOpen
  \bibfield  {author} {\bibinfo {author} {\bibfnamefont {T.~R.~T.}\
  \bibnamefont {Han}}, \bibinfo {author} {\bibfnamefont {F.}~\bibnamefont
  {Zhou}}, \bibinfo {author} {\bibfnamefont {C.~D.}\ \bibnamefont {Malliakas}},
  \bibinfo {author} {\bibfnamefont {P.~M.}\ \bibnamefont {Duxbury}}, \bibinfo
  {author} {\bibfnamefont {S.~D.}\ \bibnamefont {Mahanti}}, \bibinfo {author}
  {\bibfnamefont {M.~G.}\ \bibnamefont {Kanatzidis}},\ and\ \bibinfo {author}
  {\bibfnamefont {C.~Y.}\ \bibnamefont {Ruan}},\ }\href@noop {} {\bibfield
  {journal} {\bibinfo  {journal} {Sci. Adv.}\ }\textbf {\bibinfo {volume}
  {1}},\ \bibinfo {pages} {e1400173} (\bibinfo {year} {2015})}\BibitemShut
  {NoStop}%
\bibitem [{\citenamefont {Hellmann}\ \emph {et~al.}(2010)\citenamefont
  {Hellmann}, \citenamefont {Beye}, \citenamefont {Sohrt}, \citenamefont
  {Rohwer}, \citenamefont {Sorgenfrei}, \citenamefont {Redlin}, \citenamefont
  {Kall{\"{a}}ne}, \citenamefont {Marczynski-B{\"{u}}hlow}, \citenamefont
  {Hennies}, \citenamefont {Bauer}, \citenamefont {F{\"{o}}hlisch},
  \citenamefont {Kipp}, \citenamefont {Wurth},\ and\ \citenamefont
  {Rossnagel}}]{Hellmann2010}%
  \BibitemOpen
  \bibfield  {author} {\bibinfo {author} {\bibfnamefont {S.}~\bibnamefont
  {Hellmann}}, \bibinfo {author} {\bibfnamefont {M.}~\bibnamefont {Beye}},
  \bibinfo {author} {\bibfnamefont {C.}~\bibnamefont {Sohrt}}, \bibinfo
  {author} {\bibfnamefont {T.}~\bibnamefont {Rohwer}}, \bibinfo {author}
  {\bibfnamefont {F.}~\bibnamefont {Sorgenfrei}}, \bibinfo {author}
  {\bibfnamefont {H.}~\bibnamefont {Redlin}}, \bibinfo {author} {\bibfnamefont
  {M.}~\bibnamefont {Kall{\"{a}}ne}}, \bibinfo {author} {\bibfnamefont
  {M.}~\bibnamefont {Marczynski-B{\"{u}}hlow}}, \bibinfo {author}
  {\bibfnamefont {F.}~\bibnamefont {Hennies}}, \bibinfo {author} {\bibfnamefont
  {M.}~\bibnamefont {Bauer}}, \bibinfo {author} {\bibfnamefont
  {A.}~\bibnamefont {F{\"{o}}hlisch}}, \bibinfo {author} {\bibfnamefont
  {L.}~\bibnamefont {Kipp}}, \bibinfo {author} {\bibfnamefont {W.}~\bibnamefont
  {Wurth}},\ and\ \bibinfo {author} {\bibfnamefont {K.}~\bibnamefont
  {Rossnagel}},\ }\href@noop {} {\bibfield  {journal} {\bibinfo  {journal}
  {Phys. Rev. Lett.}\ }\textbf {\bibinfo {volume} {105}},\ \bibinfo {pages}
  {187401} (\bibinfo {year} {2010})}\BibitemShut {NoStop}%
\bibitem [{\citenamefont {Hellmann}\ \emph {et~al.}(2012)\citenamefont
  {Hellmann}, \citenamefont {Rohwer}, \citenamefont {Kall{\"{a}}ne},
  \citenamefont {Hanff}, \citenamefont {Sohrt}, \citenamefont {Stange},
  \citenamefont {Carr}, \citenamefont {Murnane}, \citenamefont {Kapteyn},
  \citenamefont {Kipp}, \citenamefont {Bauer},\ and\ \citenamefont
  {Rossnagel}}]{Hellmann2012}%
  \BibitemOpen
  \bibfield  {author} {\bibinfo {author} {\bibfnamefont {S.}~\bibnamefont
  {Hellmann}}, \bibinfo {author} {\bibfnamefont {T.}~\bibnamefont {Rohwer}},
  \bibinfo {author} {\bibfnamefont {M.}~\bibnamefont {Kall{\"{a}}ne}}, \bibinfo
  {author} {\bibfnamefont {K.}~\bibnamefont {Hanff}}, \bibinfo {author}
  {\bibfnamefont {C.}~\bibnamefont {Sohrt}}, \bibinfo {author} {\bibfnamefont
  {A.}~\bibnamefont {Stange}}, \bibinfo {author} {\bibfnamefont
  {A.}~\bibnamefont {Carr}}, \bibinfo {author} {\bibfnamefont {M.~M.}\
  \bibnamefont {Murnane}}, \bibinfo {author} {\bibfnamefont {H.~C.}\
  \bibnamefont {Kapteyn}}, \bibinfo {author} {\bibfnamefont {L.}~\bibnamefont
  {Kipp}}, \bibinfo {author} {\bibfnamefont {M.}~\bibnamefont {Bauer}},\ and\
  \bibinfo {author} {\bibfnamefont {K.}~\bibnamefont {Rossnagel}},\ }\href@noop
  {} {\bibfield  {journal} {\bibinfo  {journal} {Nat. Commun.}\ }\textbf
  {\bibinfo {volume} {3}},\ \bibinfo {pages} {1069} (\bibinfo {year}
  {2012})}\BibitemShut {NoStop}%
\bibitem [{\citenamefont {Martino}\ \emph {et~al.}(2020)\citenamefont
  {Martino}, \citenamefont {Pisoni}, \citenamefont {{\'{C}}iri{\'{c}}},
  \citenamefont {Arakcheeva}, \citenamefont {Berger}, \citenamefont {Akrap},
  \citenamefont {Putzke}, \citenamefont {Moll}, \citenamefont {Batisti{\'{c}}},
  \citenamefont {Tuti{\v{s}}}, \citenamefont {Forr{\'{o}}},\ and\ \citenamefont
  {Semeniuk}}]{Martino2020}%
  \BibitemOpen
  \bibfield  {author} {\bibinfo {author} {\bibfnamefont {E.}~\bibnamefont
  {Martino}}, \bibinfo {author} {\bibfnamefont {A.}~\bibnamefont {Pisoni}},
  \bibinfo {author} {\bibfnamefont {L.}~\bibnamefont {{\'{C}}iri{\'{c}}}},
  \bibinfo {author} {\bibfnamefont {A.}~\bibnamefont {Arakcheeva}}, \bibinfo
  {author} {\bibfnamefont {H.}~\bibnamefont {Berger}}, \bibinfo {author}
  {\bibfnamefont {A.}~\bibnamefont {Akrap}}, \bibinfo {author} {\bibfnamefont
  {C.}~\bibnamefont {Putzke}}, \bibinfo {author} {\bibfnamefont {P.~J.}\
  \bibnamefont {Moll}}, \bibinfo {author} {\bibfnamefont {I.}~\bibnamefont
  {Batisti{\'{c}}}}, \bibinfo {author} {\bibfnamefont {E.}~\bibnamefont
  {Tuti{\v{s}}}}, \bibinfo {author} {\bibfnamefont {L.}~\bibnamefont
  {Forr{\'{o}}}},\ and\ \bibinfo {author} {\bibfnamefont {K.}~\bibnamefont
  {Semeniuk}},\ }\href@noop {} {\bibfield  {journal} {\bibinfo  {journal} {npj
  2D Mater. Appl.}\ }\textbf {\bibinfo {volume} {4}},\ \bibinfo {pages} {7}
  (\bibinfo {year} {2020})}\BibitemShut {NoStop}%
\bibitem [{\citenamefont {Darancet}\ \emph {et~al.}(2014)\citenamefont
  {Darancet}, \citenamefont {Millis},\ and\ \citenamefont
  {Marianetti}}]{Darancet2014}%
  \BibitemOpen
  \bibfield  {author} {\bibinfo {author} {\bibfnamefont {P.}~\bibnamefont
  {Darancet}}, \bibinfo {author} {\bibfnamefont {A.~J.}\ \bibnamefont
  {Millis}},\ and\ \bibinfo {author} {\bibfnamefont {C.~A.}\ \bibnamefont
  {Marianetti}},\ }\href@noop {} {\bibfield  {journal} {\bibinfo  {journal}
  {Phys. Rev. B}\ }\textbf {\bibinfo {volume} {90}},\ \bibinfo {pages} {045134}
  (\bibinfo {year} {2014})}\BibitemShut {NoStop}%
\bibitem [{\citenamefont {Ritschel}\ \emph {et~al.}(2015)\citenamefont
  {Ritschel}, \citenamefont {Trinckauf}, \citenamefont {Koepernik},
  \citenamefont {B{\"{u}}chner}, \citenamefont {Zimmermann}, \citenamefont
  {Berger}, \citenamefont {Joe}, \citenamefont {Abbamonte},\ and\ \citenamefont
  {Geck}}]{Ritschel2015}%
  \BibitemOpen
  \bibfield  {author} {\bibinfo {author} {\bibfnamefont {T.}~\bibnamefont
  {Ritschel}}, \bibinfo {author} {\bibfnamefont {J.}~\bibnamefont {Trinckauf}},
  \bibinfo {author} {\bibfnamefont {K.}~\bibnamefont {Koepernik}}, \bibinfo
  {author} {\bibfnamefont {B.}~\bibnamefont {B{\"{u}}chner}}, \bibinfo {author}
  {\bibfnamefont {M.~V.}\ \bibnamefont {Zimmermann}}, \bibinfo {author}
  {\bibfnamefont {H.}~\bibnamefont {Berger}}, \bibinfo {author} {\bibfnamefont
  {Y.~I.}\ \bibnamefont {Joe}}, \bibinfo {author} {\bibfnamefont
  {P.}~\bibnamefont {Abbamonte}},\ and\ \bibinfo {author} {\bibfnamefont
  {J.}~\bibnamefont {Geck}},\ }\href@noop {} {\bibfield  {journal} {\bibinfo
  {journal} {Nat. Phys.}\ }\textbf {\bibinfo {volume} {11}},\ \bibinfo {pages}
  {328} (\bibinfo {year} {2015})}\BibitemShut {NoStop}%
\bibitem [{\citenamefont {Lee}\ \emph {et~al.}(2019)\citenamefont {Lee},
  \citenamefont {Goh},\ and\ \citenamefont {Cho}}]{Lee2019}%
  \BibitemOpen
  \bibfield  {author} {\bibinfo {author} {\bibfnamefont {S.~H.}\ \bibnamefont
  {Lee}}, \bibinfo {author} {\bibfnamefont {J.~S.}\ \bibnamefont {Goh}},\ and\
  \bibinfo {author} {\bibfnamefont {D.}~\bibnamefont {Cho}},\ }\href@noop {}
  {\bibfield  {journal} {\bibinfo  {journal} {Phys. Rev. Lett.}\ }\textbf
  {\bibinfo {volume} {122}},\ \bibinfo {pages} {106404} (\bibinfo {year}
  {2019})}\BibitemShut {NoStop}%
\bibitem [{\citenamefont {Lazar}\ \emph {et~al.}(2015)\citenamefont {Lazar},
  \citenamefont {Martincova},\ and\ \citenamefont {Otyepka}}]{Lazar2015}%
  \BibitemOpen
  \bibfield  {author} {\bibinfo {author} {\bibfnamefont {P.}~\bibnamefont
  {Lazar}}, \bibinfo {author} {\bibfnamefont {J.}~\bibnamefont {Martincova}},\
  and\ \bibinfo {author} {\bibfnamefont {M.}~\bibnamefont {Otyepka}},\
  }\href@noop {} {\bibfield  {journal} {\bibinfo  {journal} {Phys. Rev. B}\
  }\textbf {\bibinfo {volume} {92}},\ \bibinfo {pages} {224104} (\bibinfo
  {year} {2015})}\BibitemShut {NoStop}%
\bibitem [{\citenamefont {Perfetti}\ \emph {et~al.}(2005)\citenamefont
  {Perfetti}, \citenamefont {Gloor}, \citenamefont {Mila}, \citenamefont
  {Berger},\ and\ \citenamefont {Grioni}}]{Perfetti2005}%
  \BibitemOpen
  \bibfield  {author} {\bibinfo {author} {\bibfnamefont {L.}~\bibnamefont
  {Perfetti}}, \bibinfo {author} {\bibfnamefont {T.~A.}\ \bibnamefont {Gloor}},
  \bibinfo {author} {\bibfnamefont {F.}~\bibnamefont {Mila}}, \bibinfo {author}
  {\bibfnamefont {H.}~\bibnamefont {Berger}},\ and\ \bibinfo {author}
  {\bibfnamefont {M.}~\bibnamefont {Grioni}},\ }\href@noop {} {\bibfield
  {journal} {\bibinfo  {journal} {Phys. Rev. B}\ }\textbf {\bibinfo {volume}
  {71}},\ \bibinfo {pages} {153101} (\bibinfo {year} {2005})}\BibitemShut {NoStop}%
\bibitem [{\citenamefont {Perfetti}\ \emph {et~al.}(2006)\citenamefont
  {Perfetti}, \citenamefont {Loukakos}, \citenamefont {Lisowski}, \citenamefont
  {Bovensiepen}, \citenamefont {Berger}, \citenamefont {Biermann},
  \citenamefont {Cornaglia}, \citenamefont {Georges},\ and\ \citenamefont
  {Wolf}}]{Perfetti2006}%
  \BibitemOpen
  \bibfield  {author} {\bibinfo {author} {\bibfnamefont {L.}~\bibnamefont
  {Perfetti}}, \bibinfo {author} {\bibfnamefont {P.~A.}\ \bibnamefont
  {Loukakos}}, \bibinfo {author} {\bibfnamefont {M.}~\bibnamefont {Lisowski}},
  \bibinfo {author} {\bibfnamefont {U.}~\bibnamefont {Bovensiepen}}, \bibinfo
  {author} {\bibfnamefont {H.}~\bibnamefont {Berger}}, \bibinfo {author}
  {\bibfnamefont {S.}~\bibnamefont {Biermann}}, \bibinfo {author}
  {\bibfnamefont {P.~S.}\ \bibnamefont {Cornaglia}}, \bibinfo {author}
  {\bibfnamefont {A.}~\bibnamefont {Georges}},\ and\ \bibinfo {author}
  {\bibfnamefont {M.}~\bibnamefont {Wolf}},\ }\href@noop {} {\bibfield
  {journal} {\bibinfo  {journal} {Phys. Rev. Lett.}\ }\textbf {\bibinfo
  {volume} {97}},\ \bibinfo {pages} {067402} (\bibinfo {year}
  {2006})}\BibitemShut {NoStop}%
\bibitem [{\citenamefont {Yoshioka}\ \emph {et~al.}(2009)\citenamefont
  {Yoshioka}, \citenamefont {Koga},\ and\ \citenamefont
  {Kawakami}}]{Yoshioka2009}%
  \BibitemOpen
  \bibfield  {author} {\bibinfo {author} {\bibfnamefont {T.}~\bibnamefont
  {Yoshioka}}, \bibinfo {author} {\bibfnamefont {A.}~\bibnamefont {Koga}},\
  and\ \bibinfo {author} {\bibfnamefont {N.}~\bibnamefont {Kawakami}},\
  }\href@noop {} {\bibfield  {journal} {\bibinfo  {journal} {Phys. Rev. Lett.}\
  }\textbf {\bibinfo {volume} {103}},\ \bibinfo {pages} {036401} (\bibinfo
  {year} {2009})}\BibitemShut {NoStop}%
\bibitem [{\citenamefont {Vaskivskyi}\ \emph {et~al.}(2015)\citenamefont
  {Vaskivskyi}, \citenamefont {Gospodaric}, \citenamefont {Brazovskii},
  \citenamefont {Svetin}, \citenamefont {Sutar}, \citenamefont {Goreshnik},
  \citenamefont {Mihailovic}, \citenamefont {Mertelj},\ and\ \citenamefont
  {Mihailovic}}]{Vaskivskyi2015}%
  \BibitemOpen
  \bibfield  {author} {\bibinfo {author} {\bibfnamefont {I.}~\bibnamefont
  {Vaskivskyi}}, \bibinfo {author} {\bibfnamefont {J.}~\bibnamefont
  {Gospodaric}}, \bibinfo {author} {\bibfnamefont {S.}~\bibnamefont
  {Brazovskii}}, \bibinfo {author} {\bibfnamefont {D.}~\bibnamefont {Svetin}},
  \bibinfo {author} {\bibfnamefont {P.}~\bibnamefont {Sutar}}, \bibinfo
  {author} {\bibfnamefont {E.}~\bibnamefont {Goreshnik}}, \bibinfo {author}
  {\bibfnamefont {I.~A.}\ \bibnamefont {Mihailovic}}, \bibinfo {author}
  {\bibfnamefont {T.}~\bibnamefont {Mertelj}},\ and\ \bibinfo {author}
  {\bibfnamefont {D.}~\bibnamefont {Mihailovic}},\ }\href@noop {} {\bibfield
  {journal} {\bibinfo  {journal} {Sci. Adv.}\ }\textbf {\bibinfo {volume}
  {1}},\ \bibinfo {pages} {e1500168} (\bibinfo {year} {2015})}\BibitemShut
  {NoStop}%
\bibitem [{\citenamefont {Yoshida}\ \emph {et~al.}(2017)\citenamefont
  {Yoshida}, \citenamefont {Gokuden}, \citenamefont {Suzuki}, \citenamefont
  {Nakano},\ and\ \citenamefont {Iwasa}}]{Yoshida2017}%
  \BibitemOpen
  \bibfield  {author} {\bibinfo {author} {\bibfnamefont {M.}~\bibnamefont
  {Yoshida}}, \bibinfo {author} {\bibfnamefont {T.}~\bibnamefont {Gokuden}},
  \bibinfo {author} {\bibfnamefont {R.}~\bibnamefont {Suzuki}}, \bibinfo
  {author} {\bibfnamefont {M.}~\bibnamefont {Nakano}},\ and\ \bibinfo {author}
  {\bibfnamefont {Y.}~\bibnamefont {Iwasa}},\ }\href@noop {} {\bibfield
  {journal} {\bibinfo  {journal} {Phys. Rev. B}\ }\textbf {\bibinfo {volume}
  {95}},\ \bibinfo {pages} {121405(R)} (\bibinfo {year} {2017})}\BibitemShut
  {NoStop}%
\bibitem [{\citenamefont {Yu}\ \emph {et~al.}(2015)\citenamefont {Yu},
  \citenamefont {Yang}, \citenamefont {Lu}, \citenamefont {Yan}, \citenamefont
  {Cho}, \citenamefont {Ma}, \citenamefont {Niu}, \citenamefont {Kim},
  \citenamefont {Son}, \citenamefont {Feng}, \citenamefont {Li}, \citenamefont
  {Cheong}, \citenamefont {Chen},\ and\ \citenamefont {Zhang}}]{Yu2015}%
  \BibitemOpen
  \bibfield  {author} {\bibinfo {author} {\bibfnamefont {Y.}~\bibnamefont
  {Yu}}, \bibinfo {author} {\bibfnamefont {F.}~\bibnamefont {Yang}}, \bibinfo
  {author} {\bibfnamefont {X.~F.}\ \bibnamefont {Lu}}, \bibinfo {author}
  {\bibfnamefont {Y.~J.}\ \bibnamefont {Yan}}, \bibinfo {author} {\bibfnamefont
  {Y.~H.}\ \bibnamefont {Cho}}, \bibinfo {author} {\bibfnamefont
  {L.}~\bibnamefont {Ma}}, \bibinfo {author} {\bibfnamefont {X.}~\bibnamefont
  {Niu}}, \bibinfo {author} {\bibfnamefont {S.}~\bibnamefont {Kim}}, \bibinfo
  {author} {\bibfnamefont {Y.~W.}\ \bibnamefont {Son}}, \bibinfo {author}
  {\bibfnamefont {D.}~\bibnamefont {Feng}}, \bibinfo {author} {\bibfnamefont
  {S.}~\bibnamefont {Li}}, \bibinfo {author} {\bibfnamefont {S.~W.}\
  \bibnamefont {Cheong}}, \bibinfo {author} {\bibfnamefont {X.~H.}\
  \bibnamefont {Chen}},\ and\ \bibinfo {author} {\bibfnamefont
  {Y.}~\bibnamefont {Zhang}},\ }\href@noop {} {\bibfield  {journal} {\bibinfo
  {journal} {Nat. Nanotechnol.}\ }\textbf {\bibinfo {volume} {10}},\ \bibinfo
  {pages} {270} (\bibinfo {year} {2015})}\BibitemShut {NoStop}%
\bibitem [{\citenamefont {Zhang}\ \emph {et~al.}(2019)\citenamefont {Zhang},
  \citenamefont {Lian}, \citenamefont {Guan}, \citenamefont {Ma}, \citenamefont
  {Fu}, \citenamefont {Guo},\ and\ \citenamefont {Meng}}]{Zhang2019}%
  \BibitemOpen
  \bibfield  {author} {\bibinfo {author} {\bibfnamefont {J.}~\bibnamefont
  {Zhang}}, \bibinfo {author} {\bibfnamefont {C.}~\bibnamefont {Lian}},
  \bibinfo {author} {\bibfnamefont {M.}~\bibnamefont {Guan}}, \bibinfo {author}
  {\bibfnamefont {W.}~\bibnamefont {Ma}}, \bibinfo {author} {\bibfnamefont
  {H.}~\bibnamefont {Fu}}, \bibinfo {author} {\bibfnamefont {H.}~\bibnamefont
  {Guo}},\ and\ \bibinfo {author} {\bibfnamefont {S.}~\bibnamefont {Meng}},\
  }\href@noop {} {\bibfield  {journal} {\bibinfo  {journal} {Nano Lett.}\
  }\textbf {\bibinfo {volume} {19}},\ \bibinfo {pages} {6027} (\bibinfo {year}
  {2019})}\BibitemShut {NoStop}%
\bibitem [{\citenamefont {Strebel}\ and\ \citenamefont
  {Soos}(1970)}]{Strebel1970}%
  \BibitemOpen
  \bibfield  {author} {\bibinfo {author} {\bibfnamefont {P.~J.}\ \bibnamefont
  {Strebel}}\ and\ \bibinfo {author} {\bibfnamefont {Z.~G.}\ \bibnamefont
  {Soos}},\ }\href@noop {} {\bibfield  {journal} {\bibinfo  {journal} {J. Chem.
  Phys.}\ }\textbf {\bibinfo {volume} {53}},\ \bibinfo {pages} {4077} (\bibinfo
  {year} {1970})}\BibitemShut {NoStop}%
\bibitem [{\citenamefont {Chittipeddi}\ \emph {et~al.}(1987)\citenamefont
  {Chittipeddi}, \citenamefont {Cromack}, \citenamefont {Miller},\ and\
  \citenamefont {Epstein}}]{Chittipeddi1987}%
  \BibitemOpen
  \bibfield  {author} {\bibinfo {author} {\bibfnamefont {S.}~\bibnamefont
  {Chittipeddi}}, \bibinfo {author} {\bibfnamefont {K.~R.}\ \bibnamefont
  {Cromack}}, \bibinfo {author} {\bibfnamefont {J.~S.}\ \bibnamefont
  {Miller}},\ and\ \bibinfo {author} {\bibfnamefont {A.~J.}\ \bibnamefont
  {Epstein}},\ }\href@noop {} {\bibfield  {journal} {\bibinfo  {journal} {Phys.
  Rev. Lett.}\ }\textbf {\bibinfo {volume} {58}},\ \bibinfo {pages} {2695}
  (\bibinfo {year} {1987})}\BibitemShut {NoStop}%
\bibitem [{\citenamefont {Lechermann}\ \emph {et~al.}(2007)\citenamefont
  {Lechermann}, \citenamefont {Georges}, \citenamefont {Kotliar},\ and\
  \citenamefont {Parcollet}}]{Lechermann2007}%
  \BibitemOpen
  \bibfield  {author} {\bibinfo {author} {\bibfnamefont {F.}~\bibnamefont
  {Lechermann}}, \bibinfo {author} {\bibfnamefont {A.}~\bibnamefont {Georges}},
  \bibinfo {author} {\bibfnamefont {G.}~\bibnamefont {Kotliar}},\ and\ \bibinfo
  {author} {\bibfnamefont {O.}~\bibnamefont {Parcollet}},\ }\href@noop {}
  {\bibfield  {journal} {\bibinfo  {journal} {Phys. Rev. B}\ }\textbf {\bibinfo
  {volume} {76}},\ \bibinfo {pages} {155102} (\bibinfo {year}
  {2007})}\BibitemShut {NoStop}%
\bibitem [{\citenamefont {Perdew}\ \emph
  {et~al.}(1996{\natexlab{b}})\citenamefont {Perdew}, \citenamefont {Burke},\
  and\ \citenamefont {Ernzerhof}}]{Perdew1996}%
  \BibitemOpen
  \bibfield  {author} {\bibinfo {author} {\bibfnamefont {J.~P.}\ \bibnamefont
  {Perdew}}, \bibinfo {author} {\bibfnamefont {K.}~\bibnamefont {Burke}},\ and\
  \bibinfo {author} {\bibfnamefont {M.}~\bibnamefont {Ernzerhof}},\ }\href@noop
  {} {\bibfield  {journal} {\bibinfo  {journal} {Phys. Rev. Lett.}\ }\textbf
  {\bibinfo {volume} {77}},\ \bibinfo {pages} {3865} (\bibinfo {year}
  {1996}{\natexlab{b}})}\BibitemShut {NoStop}%
\bibitem [{\citenamefont {Kulik}\ and\ \citenamefont
  {Marzari}(2011)}]{Kulik2011}%
  \BibitemOpen
  \bibfield  {author} {\bibinfo {author} {\bibfnamefont {H.~J.}\ \bibnamefont
  {Kulik}}\ and\ \bibinfo {author} {\bibfnamefont {N.}~\bibnamefont
  {Marzari}},\ }\href@noop {} {\bibfield  {journal} {\bibinfo  {journal} {J.
  Chem. Phys.}\ }\textbf {\bibinfo {volume} {134}},\ \bibinfo {pages} {094103}
  (\bibinfo {year} {2011})}\BibitemShut {NoStop}%
\bibitem [{\citenamefont {Mosey}\ \emph {et~al.}(2008)\citenamefont {Mosey},
  \citenamefont {Liao},\ and\ \citenamefont {Carter}}]{Mosey2008}%
  \BibitemOpen
  \bibfield  {author} {\bibinfo {author} {\bibfnamefont {N.~J.}\ \bibnamefont
  {Mosey}}, \bibinfo {author} {\bibfnamefont {P.}~\bibnamefont {Liao}},\ and\
  \bibinfo {author} {\bibfnamefont {E.~A.}\ \bibnamefont {Carter}},\
  }\href@noop {} {\bibfield  {journal} {\bibinfo  {journal} {J. Chem. Phys.}\
  }\textbf {\bibinfo {volume} {129}},\ \bibinfo {pages} {014103} (\bibinfo
  {year} {2008})}\BibitemShut {NoStop}%
\bibitem [{\citenamefont {Mosey}\ and\ \citenamefont
  {Carter}(2007)}]{Mosey2007}%
  \BibitemOpen
  \bibfield  {author} {\bibinfo {author} {\bibfnamefont {N.~J.}\ \bibnamefont
  {Mosey}}\ and\ \bibinfo {author} {\bibfnamefont {E.~A.}\ \bibnamefont
  {Carter}},\ }\href@noop {} {\bibfield  {journal} {\bibinfo  {journal} {Phys.
  Rev. B}\ }\textbf {\bibinfo {volume} {76}},\ \bibinfo {pages} {155123}
  (\bibinfo {year} {2007})}\BibitemShut {NoStop}%
\bibitem [{\citenamefont {Tancogne-Dejean}\ and\ \citenamefont
  {Rubio}(2019)}]{Tancogne-Dejean2019}%
  \BibitemOpen
  \bibfield  {author} {\bibinfo {author} {\bibfnamefont {N.}~\bibnamefont
  {Tancogne-Dejean}}\ and\ \bibinfo {author} {\bibfnamefont {A.}~\bibnamefont
  {Rubio}},\ }\href@noop {} {\bibfield  {journal} {\bibinfo  {journal} {Phys.
  Rev. B}\ }\textbf {\bibinfo {volume} {102}},\ \bibinfo {pages} {155117}
  (\bibinfo {year} {2020})}\BibitemShut {NoStop}%
\bibitem [{SM()}]{SM}%
  \BibitemOpen
  \href@noop {} {\bibinfo {title} {See supplemental material at
  url}}\BibitemShut {NoStop}%
\bibitem [{\citenamefont {Hovden}\ \emph {et~al.}(2016)\citenamefont {Hovden},
  \citenamefont {Tsen}, \citenamefont {Liu}, \citenamefont {Savitzky},
  \citenamefont {{El Baggari}}, \citenamefont {Liu}, \citenamefont {Lu},
  \citenamefont {Sun}, \citenamefont {Kim}, \citenamefont {Pasupathy},\ and\
  \citenamefont {Kourkoutis}}]{Hovden2016}%
  \BibitemOpen
  \bibfield  {author} {\bibinfo {author} {\bibfnamefont {R.}~\bibnamefont
  {Hovden}}, \bibinfo {author} {\bibfnamefont {A.~W.}\ \bibnamefont {Tsen}},
  \bibinfo {author} {\bibfnamefont {P.}~\bibnamefont {Liu}}, \bibinfo {author}
  {\bibfnamefont {B.~H.}\ \bibnamefont {Savitzky}}, \bibinfo {author}
  {\bibfnamefont {I.}~\bibnamefont {{El Baggari}}}, \bibinfo {author}
  {\bibfnamefont {Y.}~\bibnamefont {Liu}}, \bibinfo {author} {\bibfnamefont
  {W.}~\bibnamefont {Lu}}, \bibinfo {author} {\bibfnamefont {Y.}~\bibnamefont
  {Sun}}, \bibinfo {author} {\bibfnamefont {P.}~\bibnamefont {Kim}}, \bibinfo
  {author} {\bibfnamefont {A.~N.}\ \bibnamefont {Pasupathy}},\ and\ \bibinfo
  {author} {\bibfnamefont {L.~F.}\ \bibnamefont {Kourkoutis}},\ }\href@noop {}
  {\bibfield  {journal} {\bibinfo  {journal} {Proc. Natl. Acad. Sci.}\ }\textbf
  {\bibinfo {volume} {113}},\ \bibinfo {pages} {11420} (\bibinfo {year}
  {2016})}\BibitemShut {NoStop}%
\bibitem [{\citenamefont {Medeiros}\ \emph {et~al.}(2015)\citenamefont
  {Medeiros}, \citenamefont {Tsirkin}, \citenamefont {Stafstr{\"{o}}m},\ and\
  \citenamefont {Bj{\"{o}}rk}}]{Medeiros2015}%
  \BibitemOpen
  \bibfield  {author} {\bibinfo {author} {\bibfnamefont {P.~V.~C.}\ \bibnamefont
  {Medeiros}}, \bibinfo {author} {\bibfnamefont {S.~S.}\ \bibnamefont
  {Tsirkin}}, \bibinfo {author} {\bibfnamefont {S.}~\bibnamefont
  {Stafstr{\"{o}}m}},\ and\ \bibinfo {author} {\bibfnamefont {J.}~\bibnamefont
  {Bj{\"{o}}rk}},\ }\href@noop {} {\bibfield  {journal} {\bibinfo  {journal}
  {Phys. Rev. B}\ }\textbf {\bibinfo {volume} {91}},\ \bibinfo {pages}
  {041116(R)} (\bibinfo {year} {2015})}\BibitemShut {NoStop}%
\bibitem [{\citenamefont {Ngankeu}\ \emph {et~al.}(2017)\citenamefont
  {Ngankeu}, \citenamefont {Mahatha}, \citenamefont {Guilloy}, \citenamefont
  {Bianchi}, \citenamefont {Sanders}, \citenamefont {Hanff}, \citenamefont
  {Rossnagel}, \citenamefont {Miwa}, \citenamefont {{Breth Nielsen}},
  \citenamefont {Bremholm},\ and\ \citenamefont {Hofmann}}]{Ngankeu2017}%
  \BibitemOpen
  \bibfield  {author} {\bibinfo {author} {\bibfnamefont {A.~S.}\ \bibnamefont
  {Ngankeu}}, \bibinfo {author} {\bibfnamefont {S.~K.}\ \bibnamefont
  {Mahatha}}, \bibinfo {author} {\bibfnamefont {K.}~\bibnamefont {Guilloy}},
  \bibinfo {author} {\bibfnamefont {M.}~\bibnamefont {Bianchi}}, \bibinfo
  {author} {\bibfnamefont {C.~E.}\ \bibnamefont {Sanders}}, \bibinfo {author}
  {\bibfnamefont {K.}~\bibnamefont {Hanff}}, \bibinfo {author} {\bibfnamefont
  {K.}~\bibnamefont {Rossnagel}}, \bibinfo {author} {\bibfnamefont {J.~A.}\
  \bibnamefont {Miwa}}, \bibinfo {author} {\bibfnamefont {C.}~\bibnamefont
  {{Breth Nielsen}}}, \bibinfo {author} {\bibfnamefont {M.}~\bibnamefont
  {Bremholm}},\ and\ \bibinfo {author} {\bibfnamefont {P.}~\bibnamefont
  {Hofmann}},\ }\href@noop {} {\bibfield  {journal} {\bibinfo  {journal} {Phys.
  Rev. B}\ }\textbf {\bibinfo {volume} {96}},\ \bibinfo {pages} {195147}
  (\bibinfo {year} {2017})}\BibitemShut {NoStop}%
\bibitem [{\citenamefont {Law}\ and\ \citenamefont {Lee}(2017)}]{Law2017}%
  \BibitemOpen
  \bibfield  {author} {\bibinfo {author} {\bibfnamefont {K.~T.}\ \bibnamefont
  {Law}}\ and\ \bibinfo {author} {\bibfnamefont {P.~A.}\ \bibnamefont {Lee}},\
  }\href@noop {} {\bibfield  {journal} {\bibinfo  {journal} {Proc. Natl. Acad.
  Sci. U. S. A.}\ }\textbf {\bibinfo {volume} {114}},\ \bibinfo {pages} {6996}
  (\bibinfo {year} {2017})}\BibitemShut {NoStop}%
\bibitem [{\citenamefont {Ritschel}\ \emph {et~al.}(2018)\citenamefont
  {Ritschel}, \citenamefont {Berger},\ and\ \citenamefont
  {Geck}}]{Ritschel2018}%
  \BibitemOpen
  \bibfield  {author} {\bibinfo {author} {\bibfnamefont {T.}~\bibnamefont
  {Ritschel}}, \bibinfo {author} {\bibfnamefont {H.}~\bibnamefont {Berger}},\
  and\ \bibinfo {author} {\bibfnamefont {J.}~\bibnamefont {Geck}},\ }\href@noop
  {} {\bibfield  {journal} {\bibinfo  {journal} {Phys. Rev. B}\ }\textbf
  {\bibinfo {volume} {98}},\ \bibinfo {pages} {195134} (\bibinfo {year}
  {2018})}\BibitemShut {NoStop}%
\bibitem [{\citenamefont {Butler}\ \emph {et~al.}(2020)\citenamefont {Butler},
  \citenamefont {Yoshida}, \citenamefont {Hanaguri},\ and\ \citenamefont
  {Iwasa}}]{Butler2020}%
  \BibitemOpen
  \bibfield  {author} {\bibinfo {author} {\bibfnamefont {C.~J.}\ \bibnamefont
  {Butler}}, \bibinfo {author} {\bibfnamefont {M.}~\bibnamefont {Yoshida}},
  \bibinfo {author} {\bibfnamefont {T.}~\bibnamefont {Hanaguri}},\ and\
  \bibinfo {author} {\bibfnamefont {Y.}~\bibnamefont {Iwasa}},\ }\href@noop {}
  {\bibfield  {journal} {\bibinfo  {journal} {Nat. Commun.}\ }\textbf {\bibinfo
  {volume} {11}},\ \bibinfo {pages} {2477} (\bibinfo {year}
  {2020})}\BibitemShut {NoStop}%
\bibitem [{\citenamefont {Xian}\ \emph {et~al.}(2019)\citenamefont {Xian},
  \citenamefont {Kennes}, \citenamefont {Tancogne-Dejean}, \citenamefont
  {Altarelli},\ and\ \citenamefont {Rubio}}]{Xian2019}%
  \BibitemOpen
  \bibfield  {author} {\bibinfo {author} {\bibfnamefont {L.}~\bibnamefont
  {Xian}}, \bibinfo {author} {\bibfnamefont {D.~M.}\ \bibnamefont {Kennes}},
  \bibinfo {author} {\bibfnamefont {N.}~\bibnamefont {Tancogne-Dejean}},
  \bibinfo {author} {\bibfnamefont {M.}~\bibnamefont {Altarelli}},\ and\
  \bibinfo {author} {\bibfnamefont {A.}~\bibnamefont {Rubio}},\ }\href@noop {}
  {\bibfield  {journal} {\bibinfo  {journal} {Nano Lett.}\ }\textbf {\bibinfo
  {volume} {19}},\ \bibinfo {pages} {4934} (\bibinfo {year}
  {2019})}\BibitemShut {NoStop}%
\bibitem [{\citenamefont {Giannozzi}\ \emph {et~al.}(2017)\citenamefont
  {Giannozzi} \emph {et~al.}}]{Giannozzi2017}%
  \BibitemOpen
  \bibfield  {author} {\bibinfo {author} {\bibfnamefont {P.}~\bibnamefont
  {Giannozzi}} \emph {et~al.},\ }\href@noop {} {\bibfield  {journal} {\bibinfo
  {journal} {J. Phys. Condens. Matter}\ }\textbf {\bibinfo {volume} {29}}
  (\bibinfo {year} {2017})}\BibitemShut {NoStop}%
\bibitem [{\citenamefont {Givens}\ and\ \citenamefont
  {Fredericks}(1977)}]{Givens1977}%
  \BibitemOpen
  \bibfield  {author} {\bibinfo {author} {\bibfnamefont {F.~L.}\ \bibnamefont
  {Givens}}\ and\ \bibinfo {author} {\bibfnamefont {G.~E.}\ \bibnamefont
  {Fredericks}},\ }\href@noop {} {\bibfield  {journal} {\bibinfo  {journal} {J.
  Phys. Chem. Solids}\ }\textbf {\bibinfo {volume} {38}},\ \bibinfo {pages}
  {1363} (\bibinfo {year} {1977})}\BibitemShut {NoStop}%
\end{thebibliography}
\end{document}